\documentclass{aastex}
\usepackage{emulateapj5,psfig}
\newcommand \Lya          {\hbox{Ly$\alpha$}}
\newcommand \Zsun          {\hbox{Z$_{\odot}$}}
\newcommand \Msun          {\hbox{M$_{\odot}$}}
\begin{document}
\title{A relation between Supermassive Black Hole mass and quasar metallicity?}
\author{Craig Warner\altaffilmark{1}, Fred Hamann\altaffilmark{1}, 
and Matthias Dietrich\altaffilmark{2}}
\altaffiltext{1}{Department of Astronomy, University of Florida,
211 Bryant Space Science Center, Gainesville, FL 32611-2055,
\\E-mail: warner@astro.ufl.edu, hamann@astro.ufl.edu}
\altaffiltext{2}{Department of Physics and Astronomy, Georgia State
University, Atlanta, Georgia 30303}
\begin{abstract}
	We analyze spectra for a large sample of 578 Active Galactic Nuclei
to examine the relationships between broad emission line properties and
central supermassive black hole (SMBH) mass.  We estimate SMBH masses by
applying the virial theorem to the \ion{C}{4} $\lambda 1549$ broad emission
line.  Although the FWHMs of \ion{C}{4} and $H \beta$ appear nearly unrelated
in individual objects, these FWHMs are well correlated when averaged over
sub-samples in our database.  Therefore, the lines are equally valid
indicators of the average SMBH mass in quasar samples.
Our sample spans five orders of magnitude in SMBH mass, six orders of
magnitude in luminosity, and a redshift range from $0 \leq z \leq 5$.
Most lines diminish in equivalent width with increasing black hole mass
(the usual ``Baldwin Effect")
and there are no trends with redshift.  Recent studies indicate 
that there is a relationship between SMBH mass and the overall
bulge/spheroidal component mass of the surrounding galaxy.  This relation,
together with the well-known mass-metallicity relationship among galaxies,
predicts a relationship between SMBH mass and quasar metallicity.  We
estimate the metallicity in the broad emission line region by comparing
several line ratios involving nitrogen to theoretical predictions.  We
find that the data are consistent with a trend between SMBH mass and
metallicity, with some line ratios indicating a very strong trend, but
the uncertainties in several other important line ratios are too large to
confirm or test this correlation. 
\end{abstract}

\section{Introduction}

	Supermassive black holes (SMBHs) are believed to drive the 
tremendous energy output of quasars, and more generally, active galactic
nuclei (AGNs).  Recent studies have shown that the centers of all massive
galaxies today harbor SMBHs (Gebhardt et al. 2000a; Ferrarese \& Merritt
2002).  Moreover, the mass of the black hole scales directly with the mass of
the host galaxy, or more specifically, the mass of the host galaxy's bulge
or spheroidal component.
The early results were based on a correlation between SMBH mass
and bulge luminosity, $L_{\rm bulge}$ (Magorrian et al. 1998; Laor
1998; Wandel 1999), which showed a large scatter, as much as two
orders of magnitude in SMBH mass (Ferrarese \& Merritt 2000).
Subsequent studies focused on the relationship between the masses of SMBHs
and the velocity dispersions, $\sigma$, of their host bulges or spheroids.
This relation shows considerably less scatter than the
$M_{\rm SMBH}-L_{\rm bulge}$ relation (Ferrarese \& Merritt 2000; Gebhardt et
al.  2000a; Merritt \& Ferrarese 2001; Tremaine et al. 2002).  There is also
no significant difference between the relationships measured for active
and quiescent galaxies (Gebhardt et al. 2000b; Ferrarese et al. 2001).
Recently, SMBH mass 
has also been shown to correlate strongly with the global structure of 
ellipticals and bulges (e.g. more centrally concentrated bulges have
more massive SMBHs).  This relationship is as strong as the $M_{\rm SMBH}-
\sigma$ relationship with comparable scatter (Graham et al. 2001; Erwin
et al. 2002).  Studies that carefully model the bulge light profiles of disk
galaxies and thereby obtain more accurate values of $L_{\rm bulge}$ also show
less scatter in $M_{\rm SMBH}-L_{\rm bulge}$, similar to that in the
$M_{\rm SMBH}-\sigma$ relationship (McLure \& Dunlop 2002; Erwin et al.
2002; Bettoni et al. 2003).

	The overall mass of a galaxy is also known to correlate
with its metallicity (Faber 1973; Zaritsky et al. 1994; Jablonka et al. 1996;
Trager et al. 2000).
This relationship is generally understood in terms of the gravitational
binding energy.  More massive galaxies have deeper potential wells, making
it harder for supernovae and stellar winds to blow matter out of the region.
Therefore, the gas remains in the galaxy longer, where it is further 
reprocessed by stars, leading to higher metallicities.  This mass-metallicity
relationship, combined with the relationship
between SMBH mass and galaxy mass, suggests that there should be
a relationship between SMBH mass and the metallicity of the gas clouds
surrounding quasars (Hamann \& Ferland 1993).

	The metallicity of the gas in the broad emission line region
(BLR) of quasars can be estimated by analyzing prominent emission
lines.  BLR metallicities should be representative of the gas
in the central regions of galaxies if the BLR was enriched by stars
in those environments.  Previous studies have demonstrated that we cannot
simply examine ratios of strong metal to hydrogen lines, such as
\ion{C}{4} $\lambda 1549$ / \Lya, to measure the C/H abundance.  This
is because strong collisionally excited lines like \ion{C}{4} play an
important role in cooling the gas.  Increasing the metallicity and thus
C/H leads to lower gas temperatures and an almost constant
\ion{C}{4} / \Lya\ ratio (Hamann \& Ferland 1999).  The best abundance
diagnostics for the BLR
involve ratios of lines that are i) not important in the cooling, and ii)
have similar excitation/emission requirements (Hamann et al. 2002).  Ratios
of emission lines involving nitrogen, N, are especially valuable in
determining the metallicity, $Z$, because of the expected ``secondary" N
production via the CNO cycle of nucleosynthesis in stars (Shields 1976;
Hamann \& Ferland 1992, 1993, 1999; Ferland et al. 1996; Hamann et al. 2002).
In the CNO cycle, nitrogen is produced from existing carbon and
oxygen.  The nitrogen abundance should therefore scale roughly as
N/H $\propto Z^{2}$ or N/O $\propto$ O/H $\propto Z$ (Tinsley 1980), providing
a sensitive metallicity diagnostic even when direct measures of $Z$ = O/H are
not available.  Observations of \ion{H}{2} regions indicate
that secondary nitrogen production and N/O $\propto$ O/H scaling dominate
for $Z \gtrsim 0.2-0.3$ $\Zsun$ (van Zee et al. 1998; Pettini et al. 2002).

	Some other studies suggest that there can be significant departures
from the simple N/O $\propto$ O/H relationship owing to time-dependent or
perhaps metallicity-dependent yields of N, O, and C (eg. Henry et al. 2000).
These effects are often invoked to explain the observed object-to-object
scatter in N/O versus O/H.  However, there is a general consensus that
large (solar or higher) N/O ratios indicate large (solar or higher)
metallicities.  Moreover, the most homogeneous and therefore most reliable
\ion{H}{2} region data (Pilyugin et al. 2002; Pettini et al. 2002) indicate
that the scatter in N/O decreases and the validity of N/O $\propto$ O/H
improves in the regime of roughly solar or higher metallicities.  This is
the regime of the gas near quasars,
based on studies of both their broad emission and intrinsic narrow
absorption lines (Constantin et al. 2001; Dietrich et al. 1999, 2003a; Hamann
1997; Hamann \& Ferland 1999; Hamann et al. 2002; Osmer et al. 1994;
Petitjean et al. 1994; Warner et al. 2002).  In this paper, we simply assume
that N/O $\propto$ O/H applies.  We also note that our analysis of composite
quasar spectra naturally averages over object-to-object variations.  Any
metallicity trends that we infer from the emission line ratios depend on
this assumption of N/O $\propto$ O/H in addition to measurement and
photoionization modeling uncertainties.

	We have collected spectra of over 800 ``Type 1" AGNs
(quasars and Seyfert 1 galaxies) from the HST archives and various
ground-based observing programs to examine trends in their emission line
properties (Dietrich et al. 2002).  578 of these spectra span the
rest-frame UV wavelengths needed for this study.
We compute five composite quasar spectra from this sample, representing
different intervals in SMBH mass from $10^{6}$ to $10^{10}$ \Msun.  We 
present measurements of the emission lines in these composite spectra and
investigate the relationship between SMBH mass and the metallicity of the BLR
gas.

\section{SMBH Mass Determinations}

	We estimated black hole masses by applying the virial theorem, 
$M_{\rm SMBH} = rv^{2}/G$, to the line-emitting gas, where $v$ is the
velocity dispersion and $r$ is the radial distance away from the SMBH
(Peterson \& Wandel 2000).  The relationship between $v$ and the FWHM of the
broad emission line profile depends on the kinematics and geometry, but can
be expressed as $v = k \times {\rm FWHM}$, where $k$ is a factor on the order
of unity (McLure \& Dunlop 2001; Vestergaard 2002).
Kaspi et al. (2000) express the SMBH mass as
\begin{equation}
M = 1.464 \times 10^{5} \Msun \left( {{R_{{\rm BLR}}} \over {{\rm lt-days}}} \right) \left( {{\rm FWHM}} \over {{\rm 10^{3}\ km\ s^{-1}}} \right) ^{2} \
\end{equation}
FWHM applies to the broad emission line profile, and $R_{\rm BLR}$ is
the radial distance between the BLR and the central mass/continuum source.
Equation (1) assumes the gas in the BLR is gravitationally bound and BLR
velocities are random.  The exact geometry and kinematics are still debated
(e.g. McLure \& Dunlop 2001), but are unknown, so we take this approach of
random velocities ($k = \sqrt{3}/2$) because it is the most basic and allows
for comparison with most other work.  Any offset in mass due to different
geometric or kinematic assumptions would only affect the absolute SMBH masses
and not any trends derived between metallicity and SMBH mass.
We estimate $R_{\rm BLR}$ based on
the observed relation between $R_{\rm BLR}$ for a particular line
and the continuum luminosity (Kaspi et al. 2000; Vestergaard 2002;
Wandel et al. 1999).  A particular line must be specified because
reverberation studies have shown that the BLR is radially stratified,
such that higher ionization lines tend to form closer to the central engine
than lower ionization lines (Peterson 1993).  Kaspi et al. (2000) find
\begin{equation}
R_{\rm BLR}(H \beta) = (32.9^{+2.0}_{-1.9}) \left[ {{\lambda L_{\lambda}(5100 {\rm \AA})} \over {10^{44} \ {\rm ergs \ s^{-1}}}} \right] ^{0.700 \pm 0.033} \ {\rm lt-days}
\end{equation}
where  $R_{\rm BLR}(H \beta)$ is the radial distance between the central
continuum source and the $H \beta$ emission line region and
$\lambda L_{\lambda}(5100 {\rm \AA})$ is
the continuum luminosity at 5100 \AA\ in the AGN rest frame.  We choose
to use the \ion{C}{4} $\lambda 1549$ emission line instead of $H \beta$
to estimate the SMBH mass because it is more readily observed across the entire
redshift range from $z \sim$ 0 to $z \sim$ 5.  \ion{C}{4} may in fact yield a
better estimate of SMBH mass than $H \beta$ because $H \beta$ can include a
narrow line component that forms in a different region and thus must be
subtracted before measuring the FWHM.  \ion{C}{4} has no such narrow component
(Wills et al. 1993, Vestergaard 2002).  It has recently been suggested that
\ion{Mg}{2} $\lambda 2798$ may be a better indicator of SMBH mass than
\ion{C}{4} (McLure \& Jarvis 2002).  However, \ion{C}{4} is observable over a
wider range of redshifts and is clearly the best choice among UV lines at
shorter wavelengths.  Significant blending effects are much less common for
\ion{C}{4} than other strong lines in the rest-frame UV spectrum
and absorption effects are much less common for \ion{C}{4} than for \Lya\
(Vestergaard 2002). 

	Reverberation studies indicate that the size of the broad emission
line region for \ion{C}{4} is about half that of $H \beta$ (Stirpe et al. 1994;
Korista et al.  1995; Peterson 1997; Peterson \& Wandel 1999).  We modify
Equation (2) to account for this and we use a powerlaw of the form
$F_{\nu} \propto \nu^{\alpha}$ with $\alpha = -0.4$ to estimate a conversion
between the continuum luminosity at 5100 \AA\ and 1450 \AA.  We select
$\alpha = -0.4$ based on average quasar spectra from Brotherton et al.
(2001), Vanden Berk et al. (2001), and Dietrich et al. (2002).  This yields 
\begin{equation}
R_{\rm BLR}({\rm C IV}) = 9.7 \left[ {{\lambda L_{\lambda}(1450 {\rm \AA})} \over {10^{44} \ {\rm ergs \ s^{-1}}}} \right] ^{0.7} \ {\rm lt-days}
\end{equation}

	Vestergaard (2002) finds that this technique of using the continuum
luminosity and the FWHM of \ion{C}{4} yields an estimate of the central
black hole mass that has a 1$\sigma$ uncertainty of a factor of three 
when compared to reverberation mapping studies that use $H \beta$ to estimate
SMBH masses.  From Equations (1) and (3), we obtain 
\begin{equation}
M = 1.4 \times 10^{6} \Msun \left( {{{\rm FWHM(C IV)}} \over {{\rm 10^{3}\ km\ s^{-1}}}} \right) ^{2} \left( {{\lambda L_{\lambda}(1450 {\rm \AA})} \over {{\rm 10^{44} \ {\rm ergs \ s^{-1}}}}} \right) ^{0.7} \
\end{equation}
This is essentially the same as (within 10\% of) the mass relationship
derived by Vestergaard using \ion{C}{4} and $\lambda L_{\lambda}(1350
{\rm \AA})$ (For this comparison, we again assume the powerlaw index,
$\alpha = -0.4$ between 1450 and 1350 \AA.)  The difference of $\sim10\%$ 
between the two formulae is negligible compared to the 1$\sigma$
uncertainty of a factor of three.  See also Netzer (2003) and Corbett et
al. (2003) for further discussion of the uncertainties.

	While multi-epoch spectra are preferable (to average over
variabilities), single-epoch spectra may be used if the signal-to-noise ratio
is high enough because the FWHM of single-epoch spectra are generally
consistent with the FWHMs of the mean and rms spectra obtained from
multi-epoch observations (Vestergaard 2002).  It is important to accurately
measure the FWHM because the uncertainties in SMBH mass are dominated by
the uncertainties in the FWHM.
Our composite spectra are each made up of single-epoch observations
of multiple quasars and therefore should be approximately equivalent to
multi-epoch observations of the same sources.

\section{Data \& Analysis}

	We have compiled a sample of more than 800 quasar spectra,
spanning a redshift range from $0 \lesssim z \lesssim 5$ and six orders
of magnitude in intrinsic luminosity.  The spectra were obtained by several
groups using various ground-based instruments, as well as the {\it
International Ultraviolet Explorer (IUE)} and the {\it Hubble Space Telescope
(HST)}.  One unique aspect of this sample is that it contains new observations
of faint quasars at redshift $z > 2.5$ (Dietrich et al. 2002).  Thus we
can avoid to some degree the bias toward higher luminosities at higher
redshifts that affects other magnitude-limited samples. 
578 of the spectra have UV wavelength coverage that encompasses the
range $950 \lesssim \lambda \lesssim 2050$ \AA.  Each spectrum was
transformed to the rest-frame using a redshift measured from the centroid
of the upper 50\% of the \ion{C}{4} $\lambda 1549$ profile.  See Dietrich et
al. (2002) and Dietrich et al. (2003, in prep) for more details on the AGN
sample and how luminosities were obtained.  Throughout this paper, we use the
cosmological parameters H$_{0}$ = 65 km s$^{-1}$ Mpc$^{-1}$, $\Omega_{M}$
= 0.3, and $\Omega_{\Lambda}$ = 0 (Carroll, Press, \& Turner 1992).
We used radio flux densities given in V\'{e}ron-Cetty \& V\'{e}ron (2001) to
determine the radio loudness for the quasars.

	We use an automated program to estimate the FWHM of \ion{C}{4}
in each spectrum.  The program averages over noise by applying a smoothing
routine that calculates the median of each five-pixel interval.  It then fits
a local linear continuum (contrained by the flux in
25 \AA\ wide intervals centered around 1442.5 \AA\ and 1712.5 \AA) around
\ion{C}{4}, measures the peak flux of the emission line (of the smoothed
spectrum), and uses those to estimate the FWHM of the line.  Comparisons
between the FWHMs estimated by the program and those measured manually
indicate an error of $\lesssim 10$\% in the estimates.  Lines
containing significant absorption are flagged by the program and their
FWHMs are estimated manually (by manually interpolating across the absorption
feature).  We use the FWHM of \ion{C}{4} and the
continuum luminosity, $\lambda L_{\lambda}(1450 {\rm \AA})$ to estimate the
central SMBH mass of each quasar based on Equation (4).

	We then sorted the quasars by SMBH mass into five bins: $10^{6}-10^{7}$
\Msun, $10^{7}-10^{8}$ \Msun, $10^{8}-10^{9}$ \Msun,  $10^{9}-10^{10}$
\Msun, and $>10^{10}$ \Msun, and computed five composite spectra.
Each composite spectrum is the average of all the quasar spectra in a bin
(see Table 1 for data about each composite spectrum, including the powerlaw
index, $\alpha$, of the continuum fit, the number of objects contributing to
the composite at and FWHM of \ion{C}{4} and $H \beta$, continuum luminosity,
and mean SMBH mass).  Calculating composite spectra increases the
signal-to-noise ratio significantly, making it easier to measure weaker
emission lines that may be lost in the noise in individual
spectra.

	Since narrow absorption features may influence the emission line
profiles in composite spectra, we developed a method to detect strong narrow
absorption features that exclude the contaminated spectral region of
individual spectra from the calculation of the composite spectrum.
The procedure is roughly as follows (see Dietrich et. al 2002 for more
details).  First, a preliminary mean spectrum is calculated from all of the
spectra in a given mass bin. Each individual spectrum is then divided by this
preliminary mean to derive a ratio spectrum. The ratio spectrum is smoothed
with a running boxcar function, which provides in addition to the smoothed
ratio also a root-mean-square (rms) value for each wavelength box. A narrow
absorption line will
cause a sharp increase in the rms spectrum. The contaminated spectral
regions of individual spectra, identified by these rms spikes, are excluded
from the calculation of the final composite spectra, and no interpolation
has been applied (Dietrich et al. 2002).

	We next corrected each composite spectrum for strong iron
emission lines.  To get a first estimate of the contribution of Fe emission,
we used the empirical Fe emission template that was carefully extracted from
I\,Zw\,1 by Vestergaard \& Wilkes (2001), which they very kindly provided for
this study.   The template is mostly \ion{Fe}{2} emission but contains some
\ion{Fe}{3} emission as well.  We have modified this Fe emission template int
he UV range $\lambda
\lambda 1200 - 1540$\,\AA\ in accordance with photoionization model
calculations presented by Verner et al.\,(1999). Based on our analysis of
quasar composite spectra (Dietrich et al.\,2002), we found that the strength
of Fe line emission in the wavelength range $\lambda \lambda 1420 - 1470$\,\AA\
is equal to $\sim$2 \%\ of the integrated {\it pseudo continuum} flux in
this wavelength range (Dietrich et al. 2002).  Fe contributions vary for other
wavelength ranges, but we scale the entire template to match the 2\%
contribution in the range 1420 -- 1470 \AA.  This estimate of a 2 \%
contribution is conservatively low to avoid over-subtraction of the Fe
emission template.  The spectral width of the Fe
emission features was adjusted to a width corresponding to the FWHM of the
\ion{C}{4} $\lambda 1549$ emission line profile of each quasar.  The
appropriated scaled Fe emission template was subtracted from the corresponding
composite spectrum.  In spite of its small contribution, the estimate of the
Fe line emission improves particularly the measurements of \ion{N}{3}]
$\lambda 1750$ and lines such as \ion{He}{2} $\lambda 1640$ near an
unidentified $\lambda 1600$\,\AA\ emission feature (Laor et al.\,1994;
Vestergaard \& Wilkes 2001).
 
	We used the task NFIT1D in the IRAF\footnote{The Image Reduction
and Analysis Facility (IRAF) is distributed by the National
Optical Astronomy Observatories, which is operated by the Association of
Universities for Research in Astronomy, Inc. (USRA), under cooperative
agreement with the National Science Foundation.} software package to fit
the continuum of each Fe-subtracted spectrum with a powerlaw of the
form $F_{\nu}$ $\propto$ $\nu^{\alpha}$.  
The fits were constrained by the flux in wavelength
intervals between the emission lines, namely in 25 \AA\ wide windows
centered at 1460 \AA, 1770 \AA, and 2000 \AA.  The spectral indices overall
differ from $\alpha = -0.4$ because the powerlaw fit is to a narrower
wavelength range than, for example, Dietrich et al. (2002) or Vanden Berk et
al. (2001).  However, there is a significant trend for steeper (softer) UV
spectra in the low-mass sources (see Table 1).  Figures 1 and 2 show
the final \ion{Fe}{2}-subtracted composite spectra normalized by the continuum
fits.

	To measure the broad emission lines, we use the spectral fitting
routine SPECFIT (Kriss 1994), which employs $\chi^{2}$
minimization.  We fit each line with one or more Gaussian profiles, with the
understanding that each individual Gaussian component may have no physical
meaning by itself.  Our goal in fitting the lines is simply to measure the
total line strengths free of blends.  Our strategy is to obtain the best fit
with the least number of free paramaters and, when necessary, to use the
profile of strong unblended lines, such as \ion{C}{4}, to constrain the fits
to weaker or more blended lines.  Figure 3 shows an example of these fits.
The Appendix provides details of our fitting procedure for different
lines/blends.  Table 2 lists the resulting line fluxes relative to \Lya, plus
the rest-frame equivalent
widths (REWs) as measured above the fitted continuum, and the FWHM of each
line.  The fluxes, REWs, and FWHMs given in Table 2 are from the total
fitted profiles, which can include several multiplet components and up to
three Gaussian profiles per component (as described in the Appendix).

	The primary uncertainty in our flux measurements is the continuum
location.  We estimate the 1 $\sigma$ standard deviation of our measurements
of the fluxes of \ion{C}{4} and \Lya\ to be $\lesssim$10\% based on repeated
estimates with the continuum drawn at different levels.  By the
same method, we estimate the uncertainty in \ion{N}{5}, \ion{N}{3}],
\ion{He}{2}, \ion{O}{3}], and other weaker lines to be $\sim$10--20\%.
These estimates do not include the uncertainties due to line blending, which
can be important for some of the weak lines and for \ion{N}{5} in the wing of
\Lya.  The blending and continuum placement uncertainties are both 
greater for the higher mass spectra because of the broader profiles and
generally lower REWs. 

\section{\ion{C}{4} vs. H$\beta$ as a Mass Diagnostic}

        To test the hypothesis that $R_{\rm BLR}$ for \ion{C}{4} is typically
half the value of that for $H \beta$, we compare the measured FWHM of
\ion{C}{4} to that of $H \beta$ (see Figure 4).  If the velocities are
virialized and $R_{\rm BLR}$ for \ion{C}{4} is half that of $H \beta$, we
expect the FWHM of \ion{C}{4} to be a factor of $\sqrt{2}$ larger than the
FWHM of $H \beta$.  FWHMs are estimated by an automated program as described 
in \S3.  There can be separate BLR and Narrow Line Region (NLR) contributions
to $H \beta$ (Marziani et al. 1996, Hamann et al. 1997), so each $H \beta$
line is looked at individually. In 5 out of 74 cases, where there is obvious
narrow $H \beta$ emission with the same redshift and profile as [\ion{O}{3}]
$\lambda 5007$, this narrow emission is clipped off and the FWHM of $H \beta$
is manually estimated.  There appears to be
no real correlation between the two FWHMs for individual objects, but 
averages of objects in selected ranges of \ion{C}{4} FWHM fall generally
between a 1:1 correlation and the expected $\sqrt{2}$:1 correlation (see
Figure 4).  It should be noted that observations of $H \beta$ and \ion{C}{4}
for individual objects were not necessarily made simultaneously, which may
contribute to the scatter in Figure 4.  Five composite spectra, created from
objects in selected ranges in SMBH mass (derived from \ion{C}{4} using
Eqn. 4 and plotted as open boxes with error bars), do appear to be
consistent with the expectation that FWHM(\ion{C}{4}) = $\sqrt{2}$
FWHM($H \beta$).  All spectra contributing to the composite spectra contain
\ion{C}{4}, but only some contain $H \beta$ (see Table 1).  The narrow
component of $H \beta$ was not subtracted from any spectrum before the
composite spectra were created.  In the composites, the narrow $H \beta$
contribution smoothly blends into the broad-line profile.  Filled diamonds
represent the averages of objects in selected ranges of \ion{C}{4} FWHM.
These averages contain only the individual objects (plotted with + signs in
the figure) that have data at both \ion{C}{4} and $H \beta$.  Thus, while
there is a large scatter among individual sources, the FWHMs of \ion{C}{4}
and $H \beta$ generally scale as expected in both the composite spectra and
the averages of individual objects.  Also note that any offsets in Fig. 4
affect only the absolute SMBH masses, with an uncertainty of up to a factor
of two, but do not affect any trends with SMBH mass. 

        We also estimate the SMBH mass derived from \ion{C}{4} and the
SMBH mass derived from $H \beta$.  Masses are derived from $H \beta$ using
Eqns. 1 and 2.  We compare the two masses for all 
objects in our sample that have observations of both lines (see Figure 5). 
Filled diamonds represent the averages of objects in selected ranges of
SMBH mass (as derived from \ion{C}{4}).  As in Fig. 4, these averages contain
only the individual objects that have data at both \ion{C}{4} and $H \beta$.
We find that that there is approximately a 1:1 correlation between the mass
obtained from \ion{C}{4} and that obtained from $H \beta$.  There can be
significant deviations from this for individual objects, but the relation
holds well for averages of many objects.  Hereafter, we will discuss only
the masses obtained from \ion{C}{4}.

\section{Results}

\subsection{SMBH Mass vs. Redshift, Luminosity, and FWHM}

	Figure 6 shows the redshift distribution of the entire
sample as a function of SMBH mass.  The sample was designed to span the widest
possible range of luminosity at each redshift, and the distribution of points
in Fig. 6 closely follows the distribution in $L$ vs. $z$ (see Dietrich et al.
2002).  There is no trend in SMBH mass with redshift.
We expect SMBH mass to correlate with both continuum
luminosity and FWHM because we derived our mass estimates from these
quantities (Eqn. 4).  Continuum luminosity is plotted as a function of SMBH
mass in Figure 7, and shows the expected trend with little
scatter.  This relationship is much tighter than the relationship with the
FWHM of \ion{C}{4} (see Figure 8), even though our SMBH mass estimates
depend more sensitively on FWHM than luminosity.  This may be due to the
larger dynamic range in luminosity compared to FWHM (see Netzer 2003 for
further discussion).  In all three plots (Figs. 6-8), no trend is seen based
on radio loud vs. radio quiet objects.

\subsection{Line Properties versus SMBH Mass}

	The Baldwin Effect is an observed anti-correlation between continuum
luminosity (and thus SMBH mass, since that is estimated based on luminosity)
and the equivalent widths of emission lines such as \ion{C}{4}, \Lya,
\ion{C}{3}], and \ion{O}{6} (see Baldwin 1977; Croom et al. 2002; Dietrich et
al. 2002; Espey et al. 1993; Osmer et al. 1994, Osmer \& Shields 1999;
Zheng \& Malkan 1993).  Interestingly, the
Baldwin Effect is not seen in \ion{N}{5} even though it seems to be present
to some extent in \ion{N}{3}].  Dietrich et al. (2002) discuss
the Baldwin Effect of this same sample sorted by continuum luminosity.
The Baldwin Effect is also evident in our spectra 
sorted by $M_{\rm SMBH}$.  Figure 9 plots rest-frame equivalent widths
as a function of SMBH mass (see also Figs. 1 \& 2). 
\ion{Al}{3} curiously does not show the Baldwin Effect in these composite
spectra and, if anything, its equivalent width actually increases with
SMBH mass.  This is contrary to its behavior in Dietrich et al. (2002) when
sorted by continuum luminosity.  The rest of the lines plotted in Fig. 9
show generally the same behavior as in Dietrich et al. (2002).
Dietrich et al. (2002) find that the slope of the Baldwin Effect becomes
steeper in most lines at higher luminosities.  We observe a similar trend
for steeper slopes at higher SMBH masses with the notable exception of
\ion{N}{3}], which slightly increases in REW at higher SMBH masses.

\subsection{Metallicity versus SMBH Mass}

	Figure 10 shows the metallicities inferred by comparing
emission line flux ratios for each composite spectrum to the theoretical 
results in Hamann et al. (2002).  We obtain our metallicity estimates
by comparing the emission line flux ratios to plots of metallicity vs.
line ratio based on theoretical models (see Figure 5 in Hamann et. al
2002 and Figure 3 in Warner et al. 2002).  We prefer to estimate the
metallicities based on the calculations in Hamann et al. that use a segmented
powerlaw for the photoionizing continuum shape.
This continuum shape is a good approximation to the average observed
continuum in quasars (Zheng et al. 1997; Laor et al. 1997).  It also yields
intermediate results for line ratios, such as \ion{N}{5}/\ion{He}{2}, that
are sensitive to the continuum shape (see below).

	The uncertainties in the metallicities shown in Figure 10
derive simply from the 1-$\sigma$ measurement uncertainties discussed 
in \S3. They do not reflect the theoretical uncertainties in the technique
we use to derive metallicities from the line ratios (see Hamann et
al. 2002 for discussion). Also note that there is a larger statistical
uncertainty in the lowest mass composite spectrum, because it is composed of
only nine objects (the small number of spectra in this
composite make it easier for one or two objects to dominate the composite).

        Our best estimate of the overall metallicity from each spectrum
(labeled as Average in Fig. 10) is obtained by averaging the results of
\ion{N}{3}]/\ion{C}{3}], \ion{N}{3}]/\ion{O}{3}], \ion{N}{5}/\ion{C}{4}, and
\ion{N}{5}/\ion{O}{6}.  We select these ratios because we believe they are
the most accurately measured ratios and the most reliable from a theoretical
viewpoint (Hamann et al. 2002).

	We believe the ratios involving \ion{N}{4}] are less reliable for
several reasons.  First, the \ion{N}{4}] flux is difficult to measure because
this line is weak and in the wing of the much stronger \ion{C}{4} line
profile.  Second, ratios such as \ion{N}{4}]/\ion{C}{4}, which compare a
permitted and intercombination line, are potentially more sensitive to
uncertainties in the line optical depths and radiative transfer.  Third,
perhaps as a consequence of the greater uncertainties in \ion{N}{4}], there is
more object-to-object scatter in the results derived from \ion{N}{4}]
compared to \ion{N}{3}]  (Hamann et al. 2002).  We also exclude
\ion{N}{5}/\ion{He}{2} from our estimate of the average metallicity because
i) this ratio compares a collisionally excited line (\ion{N}{5}) to a
recombination line (\ion{He}{2}), and ii) it can yield (for this continuum
shape) very high metallicities that are obtained by extrapolations beyond
the theoretical results calculated by Hamann et al. (2002). 

	All of the line ratios involving \ion{N}{3}] and \ion{N}{5}
show N/O and N/C ratios that are solar or greater.  This implies a
metallicity of $\gtrsim$ 1 \Zsun\ if N is mostly secondary.

\subsection{Do \ion{N}{3}] and \ion{N}{5} Disagree?}

        All four ratios involving \ion{N}{5} show a strong trend in
metallicity with SMBH mass.  The slope of this trend is not linear and
seems to increase as mass increases.  The ratios involving \ion{N}{3}],
however, show no apparent correlation between metallicity and SMBH mass.
We believe that this apparent discrepancy between the two ratios may be
due to the large FWHMs and small REWs of the emission lines in the composite
spectra representing higher SMBH masses.

	The spectra representing higher SMBH masses have broader lines
because the mass derivation depends on the line FWHM (Eqn. 4).  This may 
lead to systematic overestimating of the continuum if the wings of the
lines blend together and never quite reach the continuum level.  In
particular, notice the interval around 1460 \AA\ (see Figs. 1 and 2) that we
used to constrain our continuum fits (see \S3).  In the higher mass spectra,
\ion{C}{4} and \ion{Si}{4} + \ion{O}{4}] seem to blend together, never quite
reaching the true continuum, and instead forming a ``U" shape.  This effect of
overestimating the continuum could lead to underestimates of \ion{N}{3}] and
\ion{N}{4}] in
the higher mass spectra and thus negate or weaken any real trend between
SMBH mass and the flux ratios involving these lines.  For example,
we find that a drop in the continuum level of only 5\% will more than
double the measured flux in \ion{N}{3}] and \ion{N}{4}], but
increase the flux in stronger lines by only $\sim$ 15\% to 30\%.  

	In order to more thoroughly examine the effects of FWHM on the
measured line fluxes, we sort the
quasar spectra by the FWHM of \ion{C}{4} and create composite spectra for
different ranges in FWHM.  We then fit the continuum and measure the
emission line fluxes as described in \S3.  The resulting line ratios are
plotted in Figure 11.  We find that the flux ratio of \ion{C}{3}]/\ion{C}{4}
remains roughly constant at different line widths, but \ion{N}{3}]/\ion{N}{5}
declines as the lines get broader.  Because of this, we again see a
discrepancy in metallicity trends between ratios involving \ion{N}{3}] and
those involving \ion{N}{5}.  The ratios involving \ion{N}{5} indicate
increasing metallicity with increasing line width.  This is expected because
objects with broader emission lines should have higher black hole masses and
thus higher metallicities (\S1).  The \ion{N}{3}] ratios, however, show a
``U"-like behavior in Figure 11, with the narrowest lines having the highest
metallicity.  The composite spectrum with the narrowest lines is also the
only one in which the metallicities derived from \ion{N}{3}] ratios
agree with those derived from \ion{N}{5} ratios.

	There are two possible explanations for the discrepancy between
the line ratios.  The first possibility is that \ion{N}{5} is somehow
enhanced in quasars with broader emission lines (and/or larger $L$ and
$M_{\rm SMBH}$) by a process that we do not
understand.  The second possibility is that we are systematically
overestimating the continuum and thus underestimating the flux of
\ion{N}{3}] in spectra with broader lines.  There is evidence
supporting the latter hypothesis in that \ion{N}{3}]/\ion{N}{5} and
\ion{C}{3}]/\ion{C}{4} behave similarly in our overall quasar sample.
In particular, when \ion{N}{3}]/\ion{N}{5} is larger/smaller than average,
\ion{C}{3}]/\ion{C}{4} is also larger/smaller by roughly the same
factor.  The intercombination lines scale roughly together compared to
the permitted lines of the same element.  The somewhat different behaviors
of \ion{N}{3}]/\ion{N}{5} and \ion{C}{3}]/\ion{C}{4} in Figure 11
(sorted by FWHM) therefore suggest that we are underestimating the weak
\ion{N}{3}] line in the broader line sources.  The U-like shape of the
\ion{N}{3}]/\ion{O}{3}] and \ion{N}{3}]/\ion{C}{3}] points in this
figure could then be caused by an increasing underestimate of
\ion{N}{3}] as the FWHM increases from $\sim$2000 km s$^{-1}$ to
$\sim$5000 km s$^{-1}$, offset by the increasing N/O and N/C abundance
ratios for FWHM $\gtrsim 5000$ km s$^{-1}$.  Thus, we believe this is the more
likely scenario.  This scenario may also explain the fact that the
metallicities derived from all ratios agree for objects with narrow
emission lines (see also Warner et al. 2002; Baldwin et al. 2003), in
which the continuum is unlikely to be systematically overestimated.

\section{Discussion}

	Our main result from \S5 is that the data are consistent with a
trend between BLR metallicity and SMBH mass.
The hypothesis motivating this work (\S1) was that there should be
a trend between metallicity and SMBH mass with roughly the same slope as
the mass-metallicity relation in elliptical galaxies, if the BLR gas was
enriched by stars in the AGN's host galaxy.  We use a plot of ${\rm Mg_2}$
index vs. velocity dispersion, $\sigma$, in Bender et al. (1993) to estimate
the slope of the galactic mass-metallicity relation.  ${\rm Mg_2}$ index is a
line strength parameter used to derive metallicities.  Bender et al. (1993)
derive ${\rm log(Mg_2)} \propto 0.41\ {\rm log}(Z/\Zsun)$ and
${\rm log(Mg_2)} \propto 0.20\ {\rm log} (\sigma)$.  Using these
proportionalities, along with $M \propto \sigma^2$, we obtain an estimate of
the slope of the mass-metallicity relation in log space, $\Delta Z / \Delta M
\sim 0.24$.  We obtain a similar estimate ($\Delta Z / \Delta M \sim 0.34$)
using a plot of [O/H] vs. absolute blue magnitude in Zaritsky et al. (1994)
combined with a plot of ${\rm log}(L_{B})$ vs. ${\rm log}(M)$ in van Albada
et al. (1995).  We also derive ${\rm log}(Z) \propto 0.38 \pm 0.07\
{\rm log}(M_{\rm SMBH})$ from the $[Z/{\rm H}] \propto 0.76 \pm 0.13\
{\rm log}(\sigma)$ relation given in Trager et al. (2000).
We employ a $\chi^{2}$ minimization routine to obtain
linear fits to the plots of metallicity vs. SMBH mass shown in Figure 10.
These fits yield a slope of $\Delta Z / \Delta M \sim 0.2 - 0.3$ for
all ratios involving \ion{N}{5}.  This is consistent with the slope
of the mass-metallicity relation in elliptical galaxies as expected.
The slope for the average metallicity vs. SMBH mass is $\sim$0.2,
slightly lower than the slopes for the \ion{N}{5} ratios.  The slope for
ratios involving \ion{N}{3}] is $\sim$0, but the slope from only the
three highest mass spectra for these ratios is $\sim 0.1 - 0.2$.  This
may be a lower bound on the slope of metallicity vs. SMBH mass because
in the highest mass spectra, the increasing N/O and N/C abundance ratios
are starting to dominate the underestimation of \ion{N}{3}] caused by
the systematic overestimation of the continuum in spectra with broad lines.

	It should be noted that the quasar-BLR metallicities are several
times higher than the galactic values derived in the studies listed above.
This apparent discrepancy can be understood in the context of normal galaxy
evolution. Quasar BLRs sample an earlier evolutionary stage and frequently a
different physical location in galaxies compared to the \ion{H}{2} regions or
average spectra of stars in present day galaxies (Hamann \& Ferland 1999).
However, the similar slope of the mass-metallicity relationship in both
quasars and galaxies supports the premise that the gas in the BLR was
processed/enriched by the surrounding stellar population.

	The evidence presented here for a mass-metallicity relationship
among quasars does not prove that SMBH mass is the fundamental parameter
affecting BLR metallicity.  However, in these composite spectra, sorted by
SMBH mass, the slope of the trend between {\it luminosity} and metallicity
is at least as steep as (and possibly marginally steeper than) the slope is
for composite spectra sorted by luminosity (see Dietrich et al. 2002;
Dietrich et al. 2003, in prep).  Also, composite spectra created from
different ranges in Eddington ratio, $L_{bol}/L_{edd}$, show a slight trend
with luminosity, but no trend with either SMBH mass or metallicity (see Warner
et al. 2003, in prep).  These results suggest that SMBH mass could be the
fundamental parameter.  Further support for mass as the fundamental parameter
comes from the similarity of the mass-metallicity relationship among quasars
to the corresponding relationship among galaxies.  These ideas are also
consistent with the growing evidence (e.g. Shields et al. 2002) for a close
relationship between quasars and their host galaxies.  In particular, the
relationship between SMBH mass and host galaxy mass (see refs. in \S1)
indicates that i) SMBHs are a natural byproduct of galaxy formation, and ii)
essentially all massive galaxies were at one time ``active."  Quasars at high
redshifts are signposts of the last stages in the formation of these massive
SMBHs in the cores of young galactic spheroids.  The high metallicities near
these objects reported here and elsewhere (Constantin et al. 2001; Dietrich et
al.  1999, 2003; Hamann 1997; Hamann \& Ferland 1999; Hamann et al. 2002;
Osmer et al. 1994; Petitjean et al. 1994; Warner et al. 2002) imply that the
gas has {\it already} (at the quasar epoch) been substantially
processed/enriched by multiple generations of massive stars.  This rapid
evolution at high redshifts, though, is well within the parameters
derived in some recent simulations, which show that the densest protogalactic
condensations can form stars and reach solar or higher metallicities at
$z \gtrsim 6$ (Gnedin \& Ostriker 1997; Haiman \& Loeb 2001).

\section{Summary \& Conclusions}

	We have investigated a large sample of 578 AGNs for a trend between
metallicity and SMBH mass.  The sample covers a redshift range from
$0 \lesssim z \lesssim 5$, six orders of magnitude in luminosity, and five
orders of magnitude in SMBH mass.  We estimate SMBH masses using the virial
theorem and formulae given in Kaspi et al. (2000).  To improve the
signal-to-noise ratio and average over object-to-object variations, we
produce composite spectra representing each decade in SMBH mass.  Composite
spectra allow us to better measure weak emission lines and minimize the
influence of irregularities in any individual quasar.  After a powerlaw
continuum fit, multi-component Gaussian profiles are fit to emission lines
to measure their fluxes.  Metallicities are then estimated by comparing
emission line ratios involving nitrogen to theoretical predictions in
Hamann et al. (2002).  Our main results are as follows.

        1) We estimate SMBH masses based on both \ion{C}{4} and $H \beta$ and
find approximately a 1:1 correlation for averages of many objects (Fig. 5).
We conclude that it is valid to use \ion{C}{4} instead of $H \beta$ to
estimate average black hole masses in samples of AGNs.

        2) There is no trend between SMBH mass and redshift, but as expected,
SMBH mass shows a positive correlation with both luminosity and the FWHM
of \ion{C}{4}.

        3) We observe the usual Baldwin Effect in composite spectra
representing different ranges in SMBH mass.  In agreement with other
studies (see Dietrich et al. 2002), \ion{N}{5} does not show the Baldwin
Effect even though it seems to be present to some extent in \ion{N}{3}].
However, \ion{Al}{3} curiously does not exhibit the Baldwin Effect either,
and if anything, its REW actually increases with increasing SMBH mass.

        4) We find a trend in metallicity with SMBH mass for emission line
ratios involving \ion{N}{5}, but not for those involving \ion{N}{3}] or
\ion{N}{4}].  This is consistent with the findings of Dietrich et al.
(2002).  They find a lack of a Baldwin Effect for \ion{N}{5}, which in the
present study leads to emission line ratios involving \ion{N}{5} increasing
with increasing luminosity.  However, Dietrich et al. (2003a, 2003b) find that
metallicities derived from ratios involving both \ion{N}{3}] and \ion{N}{5}
yield results of solar or greater, for samples of 70 high redshift ($3.9
\lesssim z \lesssim 5.0$) quasars.

        5) We conclude that the data are consistent with a trend between
SMBH mass and metallicity and some of the data (\ion{N}{5} ratios)
indicate that there is a very strong trend.  \ion{N}{3}] seems to indicate
no trend; however, upon further examination, we believe that the
uncertainties in \ion{N}{3}] and \ion{N}{4}] are too large to confirm or
test any trend.  We note that the correlation between SMBH mass and FWHM may
lead to systematic overestimates of the continuum and thus underestimates of
\ion{N}{3}] and \ion{N}{4}] in the higher mass spectra.

	6) We estimate the slope of the galactic mass-metallicity relationship
in log space to be $\Delta Z / \Delta M \sim 0.2 - 0.3$ (see \S6).  This
slope is consistent with linear fits to the plots of metallicity vs. SMBH
mass in Fig. 10 that are derived from ratios involving \ion{N}{5}, as well
as the plot of ``average" metallicity vs. SMBH mass.  The similarity in the
slope of the mass-metallicity relationship in both quasars and galaxies
supports the premise that the gas in the BLR was processed/enriched by
the surrounding stellar environment and contributes to the growing evidence
for a close relationship between quasars and their host galaxies. 

\noindent {\it Acknowledgements:} We are very grateful to Marianne Vestergaard
for providing the UV Fe emission template for this study and to Fred Chaffee,
Anca Constantin, Craig Foltz, Vesa Junkkarinen, and Joe Shields for their
direct participation in reducing or acquiring some of the ground-based
spectra.  We acknowledge financial support from the NSF
via grant AST99-84040 and NASA via grant NAG5-3234. 

\appendix
	We fit \ion{C}{4} $\lambda 1549$ with three Gaussian profiles. Three
Gaussians are necessary to obtain a good fit because the emission lines
are not Gaussian by nature and contain broad wings.  We ignore the doublet
splitting in \ion{C}{4} because it is small compared to the observed line
widths.  \ion{N}{4}] $\lambda 1486$, where it is measureable, is fit with
a \ion{C}{4} profile.  That is, we fit it with three Gaussian profiles
that have their FWHMs tied to the values fit to \ion{C}{4} and their
relative fluxes and relative wavelength shifts are tied to the same ratios
as in the fit to \ion{C}{4}.  For the two highest mass composite spectra,
we fit only an upper-bound to \ion{N}{4}] because it is too blended with
the wing of \ion{C}{4} to measure.  \ion{He}{2} $\lambda 1640$ is fit with
two Gaussian profiles, one broad and one narrow and \ion{O}{3}]
$\lambda 1665$ is fit with a \ion{C}{4} profile.  The weak and badly blended
doublet \ion{Si}{2} $\lambda \lambda 1527, 1533$ is fit with one Gaussian
profile for each component.

	${\rm \Lya}$ $\lambda 1216$, like \ion{C}{4}, is fit with three
Gaussian profiles, but only the red side of the line is used for $\chi^{2}$
minimization because of contamination by the \Lya\ forest on the blue side.
Each component of the \ion{N}{5} $\lambda 1240$ doublet is fit with a
\ion{C}{4} profile because \ion{N}{5} is strongly blended with the wing
of \Lya.  This connection between \ion{C}{4} and \ion{N}{5} can be
justified because they are both high ionization lines with similar
excitation properties (Ferland et al. 1998).  The relative fluxes of the
two components of the \ion{N}{5} doublet are set to the ratio of 
the statistical weight, $g$, times the $A$ value for each transition
(approximately 2:1 in this case), because this yields a better fit than
using a 1:1 ratio.  We use atomic data obtained from the National Institute
of Standards and Technology
(http://aeldata.phy.nist.gov/PhysRefData/contents-atomic.html) to calculate
$g$ times the $A$ value for each transition.  The weak and blended
multiplet \ion{Si}{2} $\lambda \lambda 1260, 1264, 1265$ is fit with one
Gaussian profile for each component.  

	We fit \ion{O}{1} $\lambda 1303$ with a single Gaussian profile
and \ion{C}{2} $\lambda 1335$ with two Gaussians, one for each component
of the doublet.  The fluxes of the two \ion{C}{2} components are tied to
a 1:1 ratio because this yields a better fit than a 2:1 ratio.  The FWHMs 
and relative wavelengths are also tied together.

	We fit the doublet \ion{Si}{4} $\lambda \lambda 1393, 1403$ with
one Gaussian profile for each of the two components.  The fluxes of these
components are tied to a 1:1 ratio because this yields a better fit than
a 2:1 ratio.  \ion{O}{4}] $\lambda 1403$, a blended multiplet of five
lines, is fit with one Gaussian profile for each of its five components.
The flux ratios of the components in this case are tied to the ratio of
$g$ times the $A$ value for each transition using atomic data obtained
from Nussbaumer \& Storey (1982).  Because \ion{Si}{4} and \ion{O}{4}]
are badly blended, we tie the FWHMs and relative wavelengths together
for all seven components of the \ion{Si}{4} - \ion{O}{4}] blend.  We list
the sum of the blend in Table 2.

	We fit the \ion{N}{3}] $\lambda 1750$ multiplet with five
\ion{C}{4} profiles, one for each component of the multiplet.  We use
\ion{C}{4} profiles to constrain the fit because \ion{N}{3}] is a weak
intercombination line.  The relative fluxes of the five components are
tied to the ratio of $g$ times the $A$ value for each transition because
this yields the best fit.  The FWHMs and relative wavelengths are all
tied together.

	We use two Gaussian profiles to fit \ion{C}{3}] $\lambda 1909$
because two Gaussians provide a fit that is just as good as the fit
obtained by using three Gaussians.  The \ion{Al}{3} $\lambda \lambda
1855, 1863$ doublet is fit with one Gaussian profile for each component.
The flux ratio is tied to the ratio of $g$ times the $A$ value for each
transition and the FWHMs and relative wavelengths of both components are
tied together.  \ion{Si}{3}] $\lambda 1892$ is fit with one Gaussian profile
and has its FWHM and relative wavelength tied to \ion{Al}{3} because it
is badly blended with \ion{C}{3}]. 

	We fit a local continuum around \ion{O}{6} $\lambda 1035$, which was
constrained by the flux in wavelength intervals centered at 960 \AA\ and
1100 \AA.  We then fit each component of the \ion{O}{6} $\lambda 1035$
doublet with two Gaussian profiles.  The fluxes are tied to a 3:2 ratio,
compromising between 1:1 and 2:1, because this yields the best fit.  Where
they are measureable, \ion{C}{3} $\lambda 977$ and \ion{N}{3} $\lambda 990$
are fit with one Gaussian profile each, and their FWHMs and relative
wavelengths are tied together.  When calculating fluxes from measured REWs,
we scale \ion{O}{6} as if it were sitting on top of a segmented powerlaw
continuum, defined by our previous continuum fit redward of 1250 \AA\ and
by a powerlaw with index $\alpha = -1.76$ (Telfer et al. 2002) blueward of
1250 \AA.

	Each of the five spectra is fit in this fashion, with the following
exceptions for the broadest ($M \sim 10^{10}$ \Msun) spectrum, which has
many extremely blended lines.  In this spectrum, we tie the relative
wavelengths of the three Gaussian profiles used to fit \ion{C}{4} and \Lya\
and the two Gaussian profiles used to fit \ion{C}{3}] to the corresponding
ratios from our fit to the $M \sim 10^{9}$ \Msun\ sample.  We do not
attempt to fit \ion{N}{4}] or \ion{Si}{2} $\lambda \lambda 1527, 1533$,
as they are undetectable due to the broad wing of \ion{C}{4}.  It should
also be noted that in our fit to this spectrum, the broader Gaussian profile
of our fit to \ion{He}{2} is blueshifted by 15 \AA.

\newpage
\begin{figure}
\vbox{
\centerline{
\psfig{figure=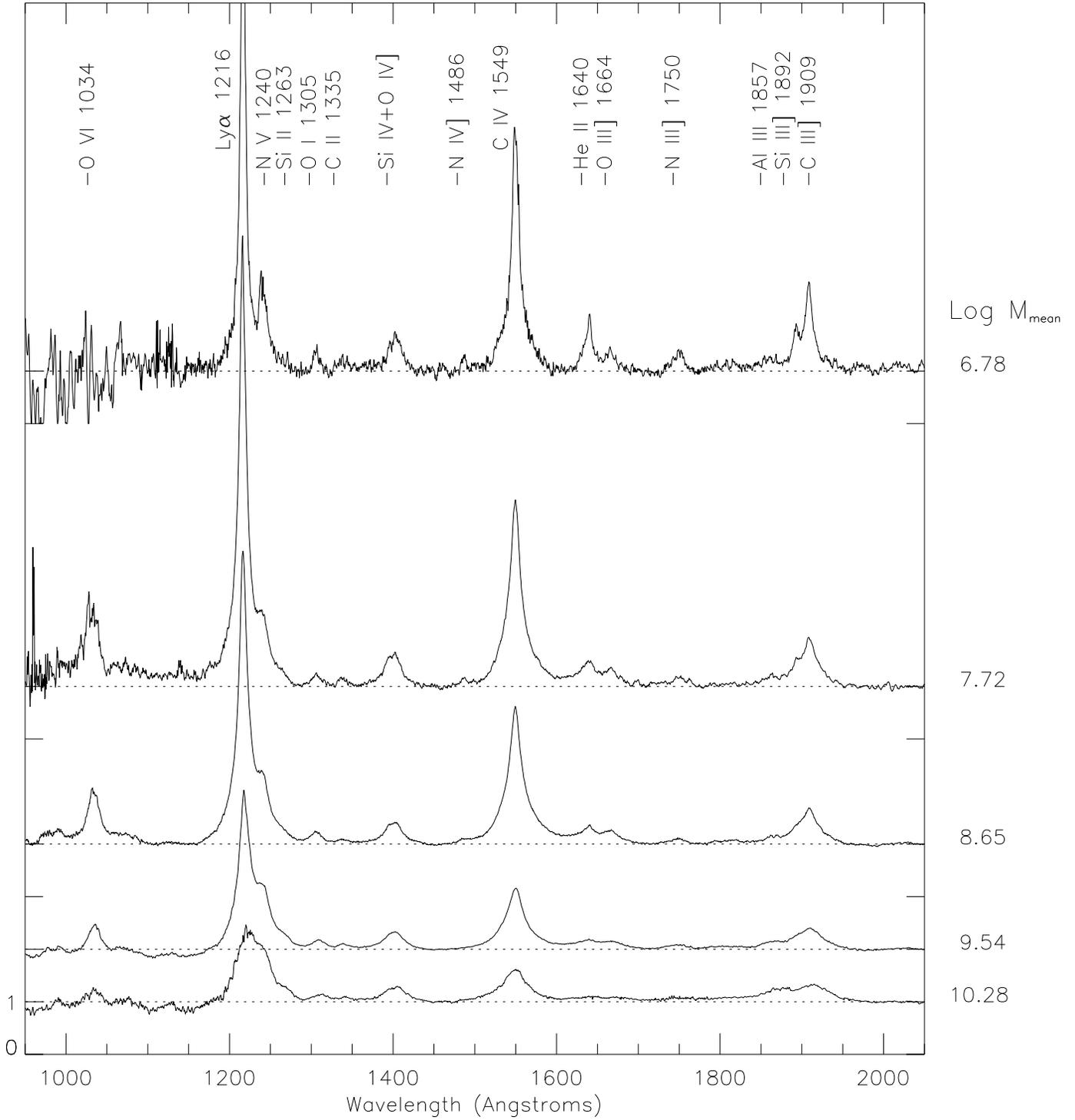}
}
\vskip 14pt
\caption
{
Normalized composite spectra are shown for all five SMBH mass bins.  The
spectra were normalized by a power law continuum fit.  The horizontal dashed
lines indicate the continuum level for the individual normalized composite
spectra, which were vertically shifted for better display.  The normalized
continuum strength is shown for the spectrum at the bottom of the figure.
This same vertical scale applies to all other spectra, although the tick marks
for '0' and '1' are not labelled.  The mean SMBH mass of each
spectra is listed to the right.  The Baldwin Effect is clearly seen in
these spectra.}
}
\end{figure}

\begin{figure}
\vbox{
\centerline{
\psfig{figure=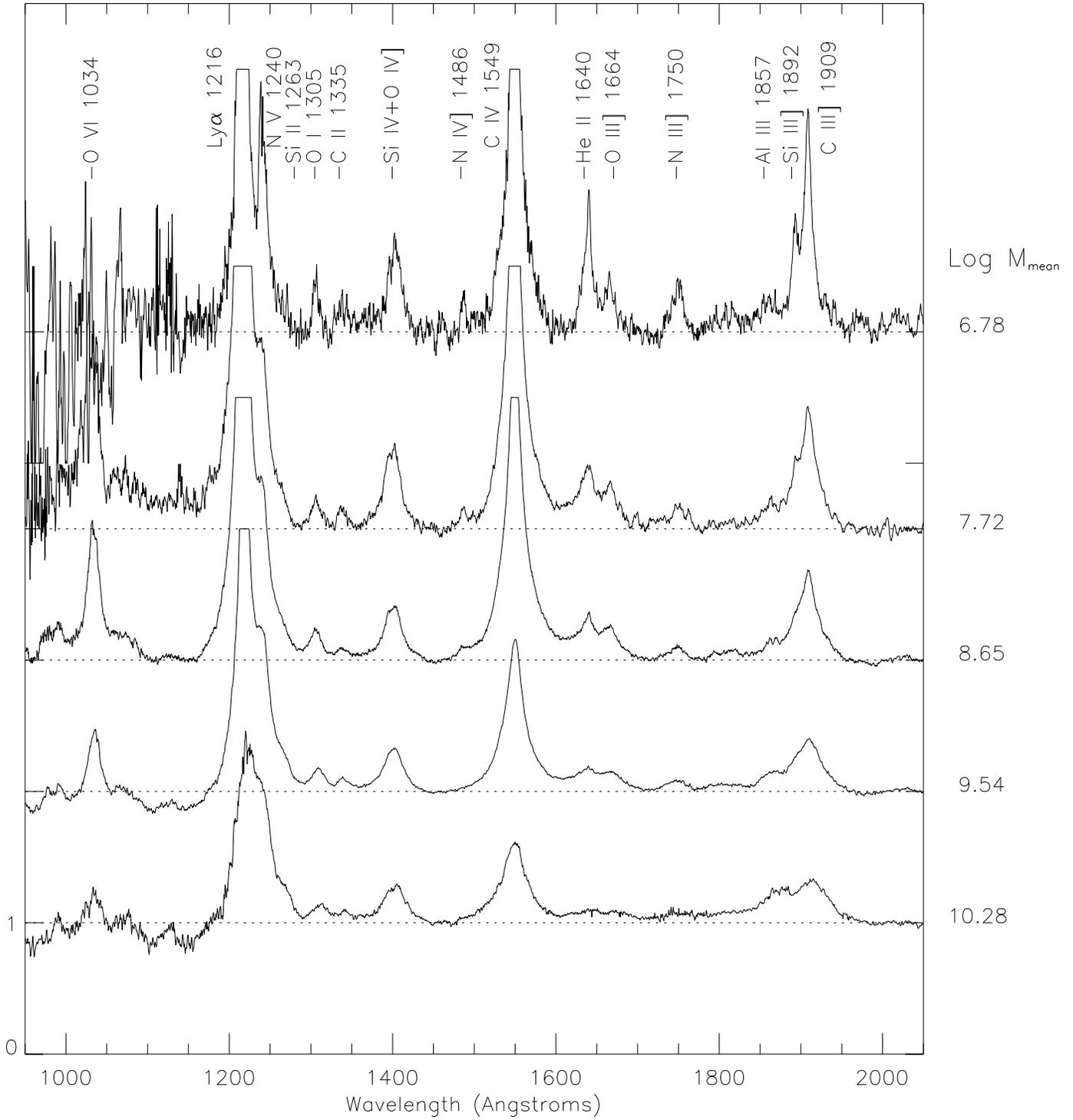}
}
\vskip 14pt
\caption
{
Same as Fig. 1, but with an expanded vertical scale to display the dependence
of the relative strength of weaker emission lines as a function of SMBH
mass.  Strong emission lines with flat tops are truncated for easier display.}
}
\end{figure}

\begin{figure}
\vbox{
\centerline{
\psfig{figure=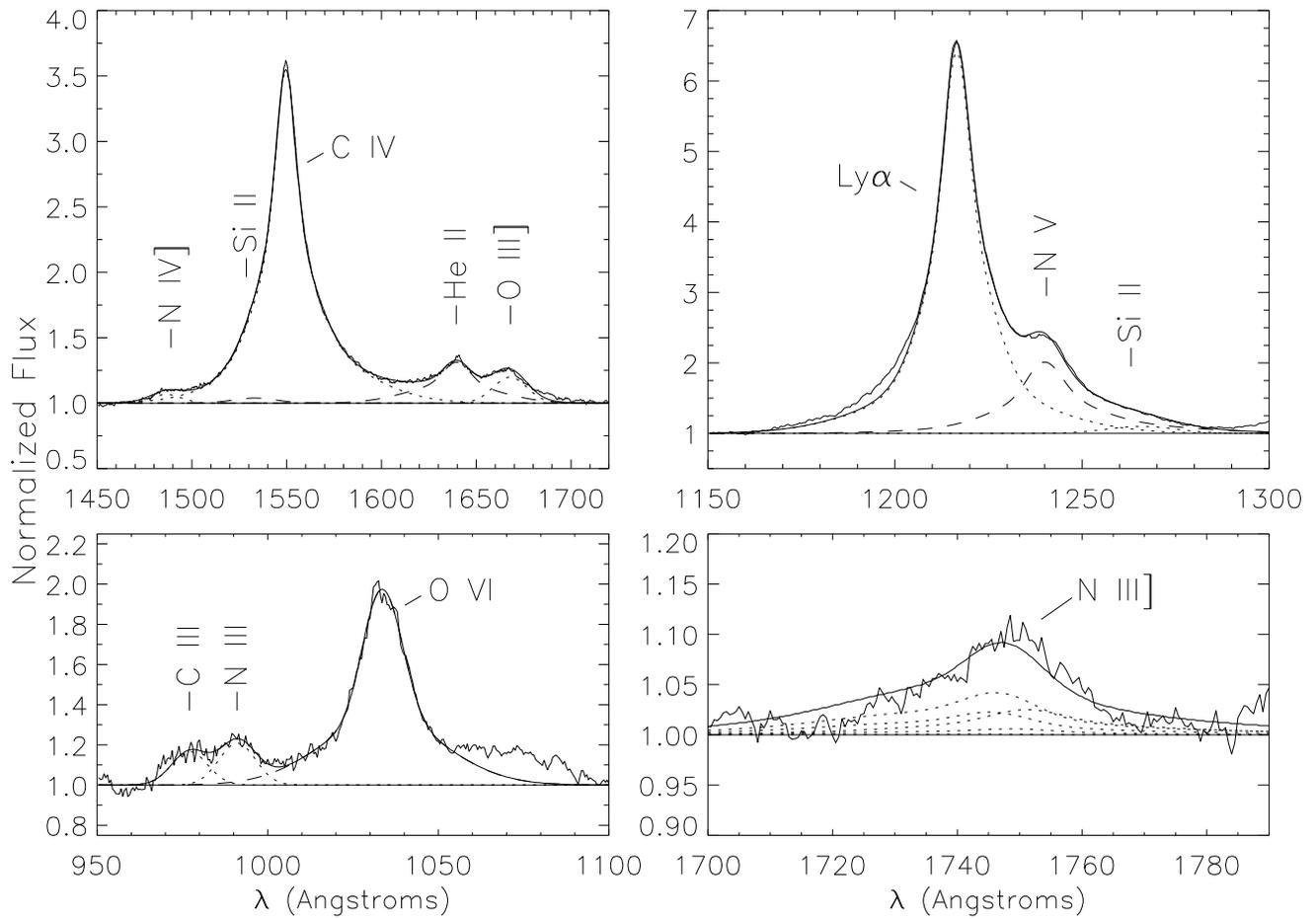}
}
\vskip 14pt
\caption
{
Multi-component Gaussian fits (smooth solid curves) to observed emission
lines (jagged solid curves) for a composite spectrum of $M\sim10^{8}$ \Msun.
The continuum is normalized by a power law
continuum fit.  The continuum (at unity) and composite line fits are
plotted.  The individual Gaussian components of the fits are shown for the fit
to \ion{N}{3}] as dotted lines.  In the other panels, the dashed and dotted
lines represent the sum contribution of each labelled emission line.  There
is an unidentified bump at $\sim$1070 \AA\ that is not attributed to
\ion{O}{6} (see Hamann et al. 1998).}
}
\end{figure}

\begin{figure}
\vbox{
\centerline{
\psfig{figure=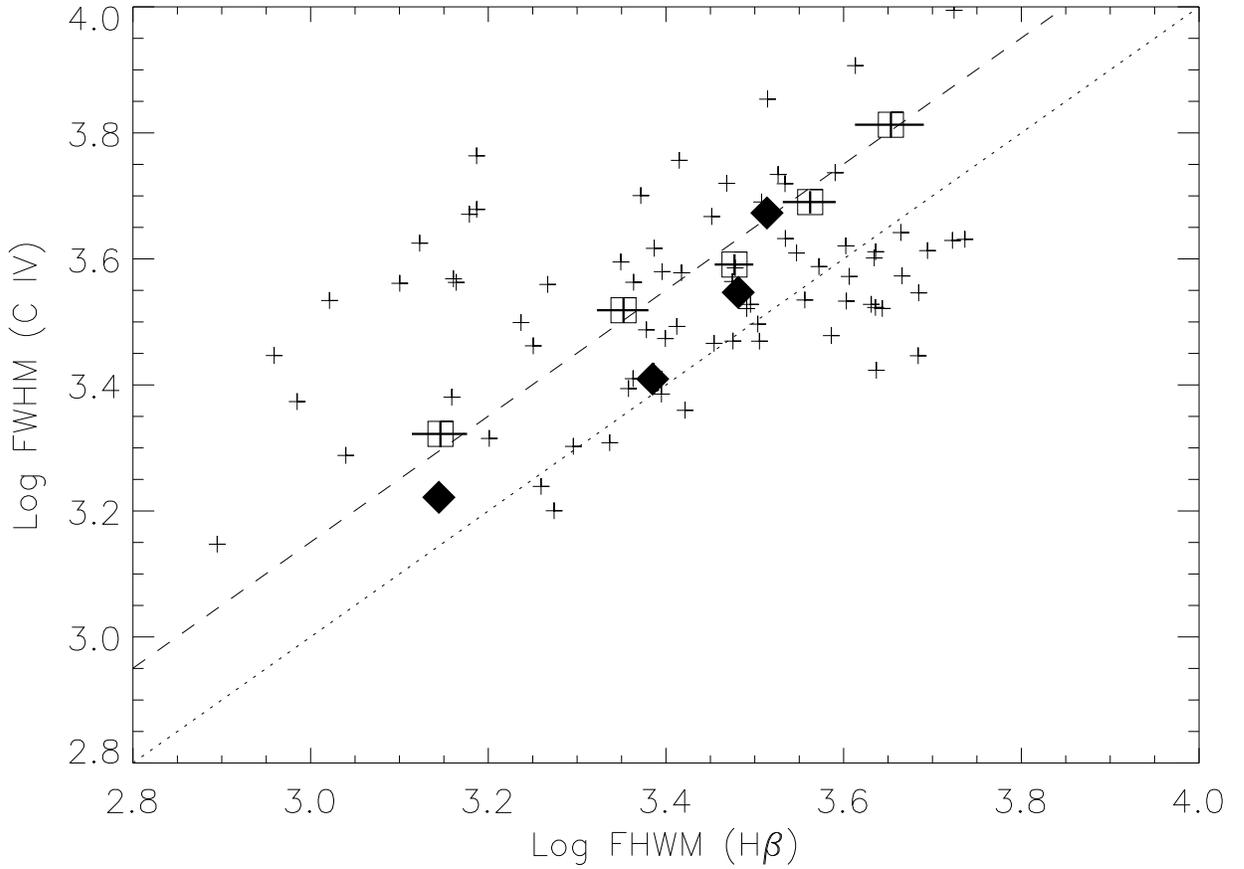}
}
\vskip 14pt
\vskip 14pt
\caption{
Comparison of the measured FWHM of \ion{C}{4} to that of $H \beta$ for all
objects in our sample with observations of both lines.  In five cases,
obvious narrow
$H \beta$ emission was subtracted prior to measuring its FWHM.  The open
boxes represent five composite spectra created from objects in
selected ranges of SMBH mass.  The error bars drawn on the boxes represent
the $\pm$ 1$\sigma$ ranges for the FWHMs of \ion{C}{4} and $H \beta$ for 
these composite spectra.  The filled diamonds represent averages of
objects in selected ranges of \ion{C}{4} FWHM ($<$ 2000 km/s, 2000--3000 km/s,
3000--4000 km/s, and 4000--6000 km/s).  The dotted line is a 1:1
ratio and the dashed line is the expected ratio of FWHM(\ion{C}{4}) $=
\sqrt{2}$ FWHM($H \beta$).}
}
\end{figure}

\begin{figure}
\vbox{
\centerline{
\psfig{figure=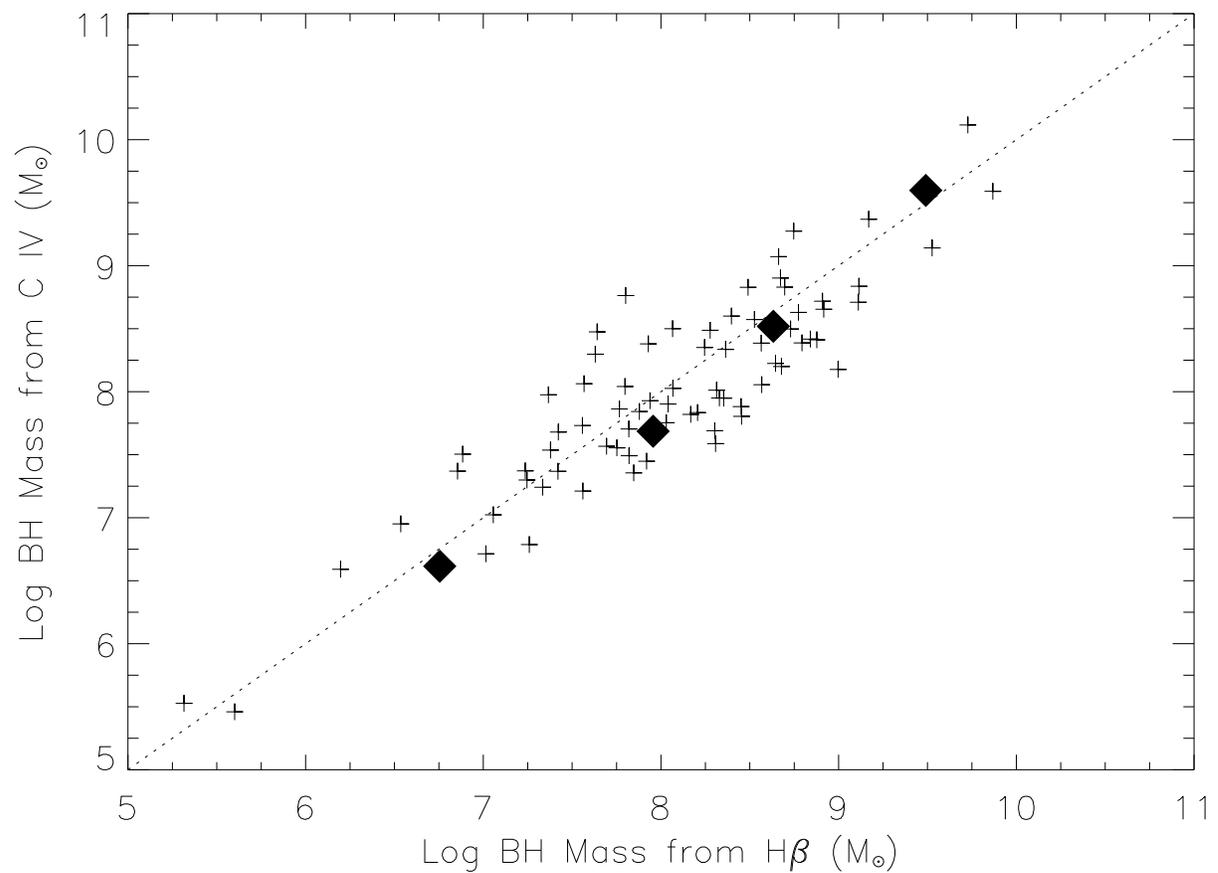}
}
\vskip 14pt
\vskip 14pt
\caption{
Comparison of the SMBH mass derived from \ion{C}{4} to that derived from
$H \beta$ for all objects in our sample with observations of both lines.
The filled diamonds represent averages of objects in selected SMBH mass
ranges.  The dotted line is a 1:1 ratio (the RMS scatter about the 1:1
relationship is 0.39 for the individual objects and 0.17 for the averages).
For composite spectra of many objects, both methods yield approximately the
same mass.}
}
\end{figure}

\begin{figure}
\vbox{
\centerline{
\psfig{figure=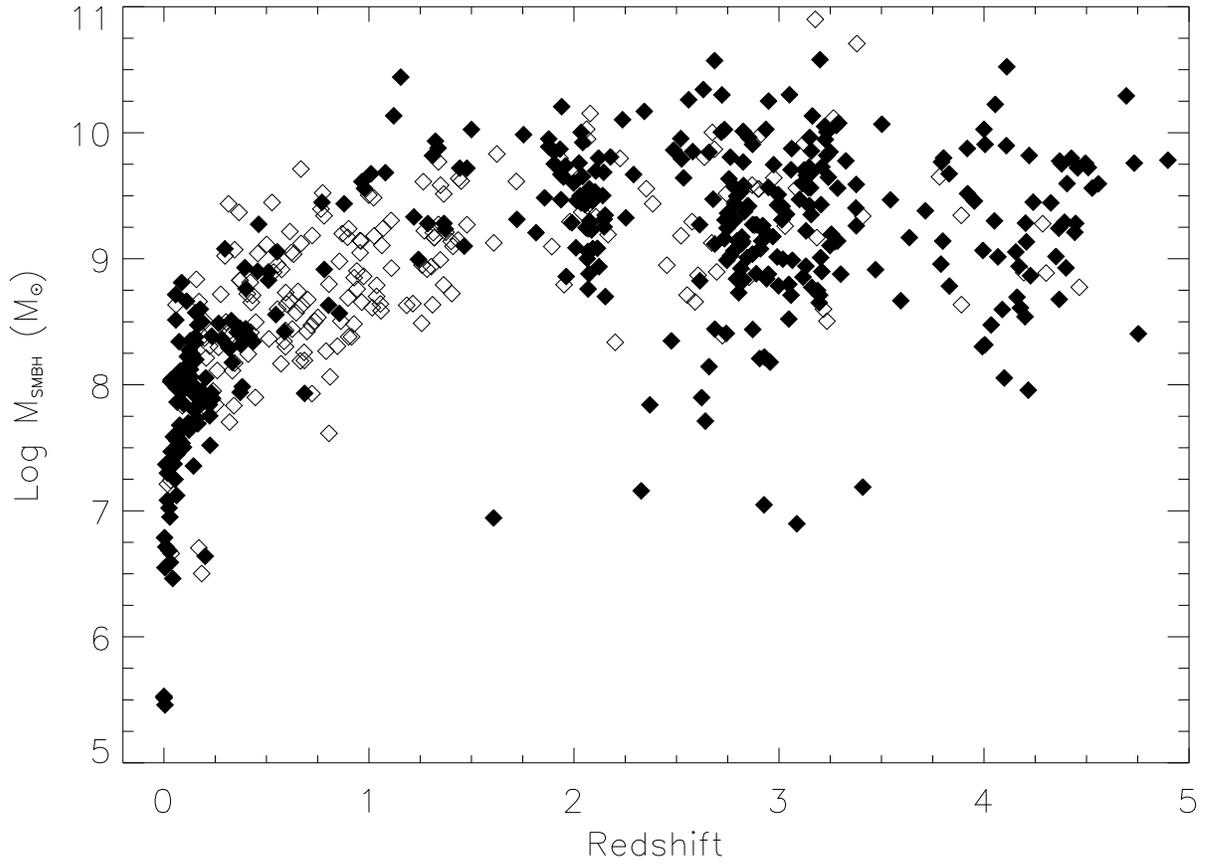}
}
\vskip 10pt
\caption{
Redshift distribution of the AGN sample as a function of SMBH mass.
The filled diamonds represent radio-quiet quasars and the open diamonds
represent radio-loud quasars.  There is clearly no trend in SMBH mass with
redshift (for $z \gtrsim 0.3$).}
}
\end{figure}

\begin{figure}
\vbox{
\centerline{
\psfig{figure=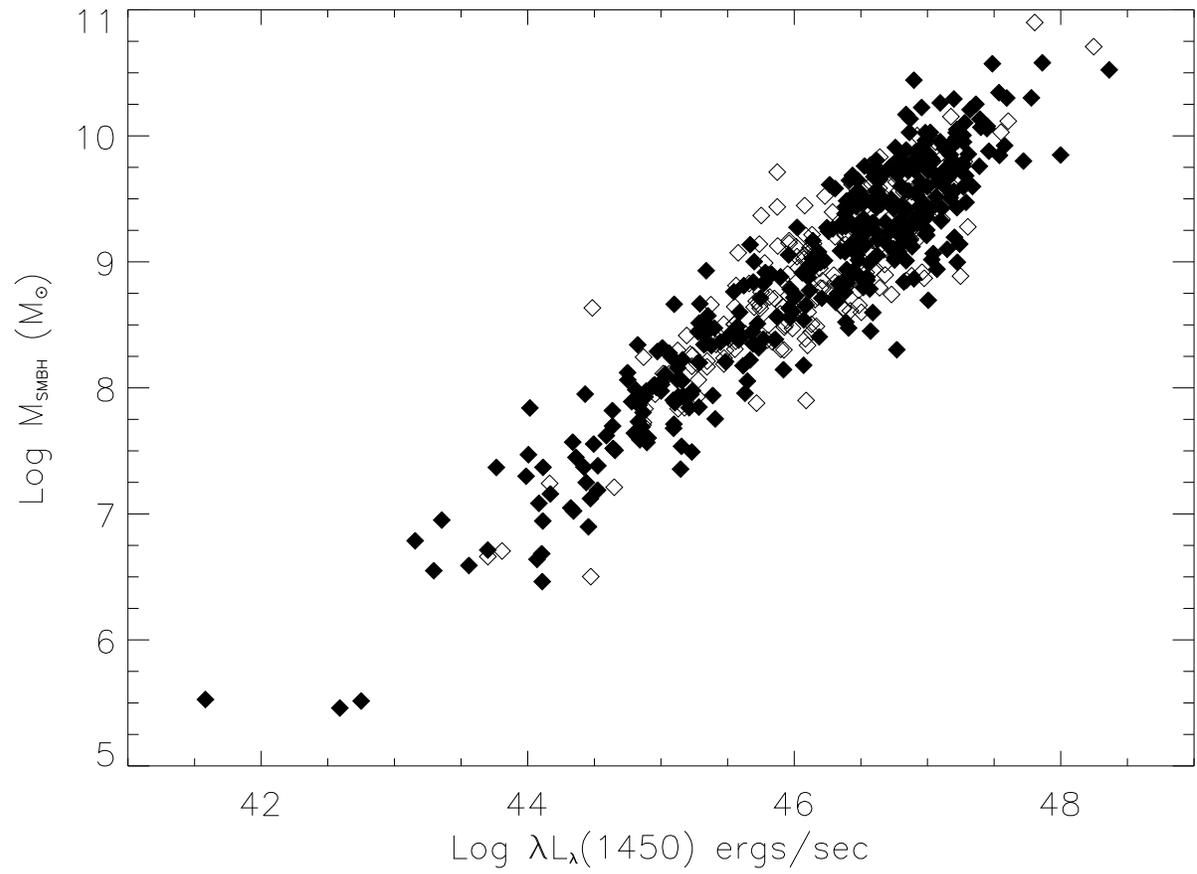}
}
\vskip 14pt
\caption
{
Continuum luminosity versus SMBH mass.  There is a very strong correlation
with little scatter.  Symbols as in Fig. 6.}
}
\end{figure}

\begin{figure}
\vbox{
\centerline{
\psfig{figure=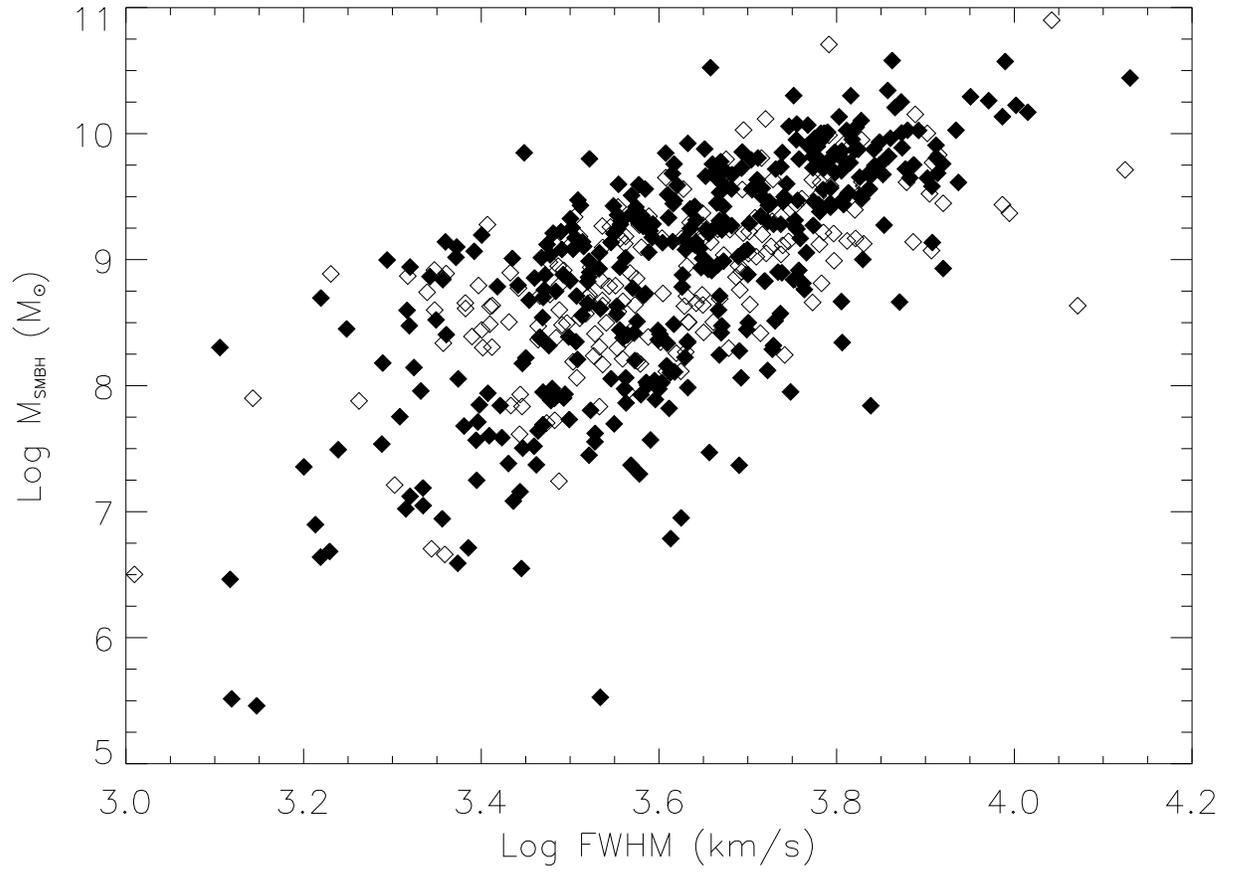}
}
\vskip 10pt
\caption{
The FWHM of \ion{C}{4} versus SMBH mass.  A trend is
seen, but with a large amount of scatter.  Symbols as in Fig. 6.}
}
\end{figure}

\begin{figure}
\vbox{
\centerline{
\psfig{figure=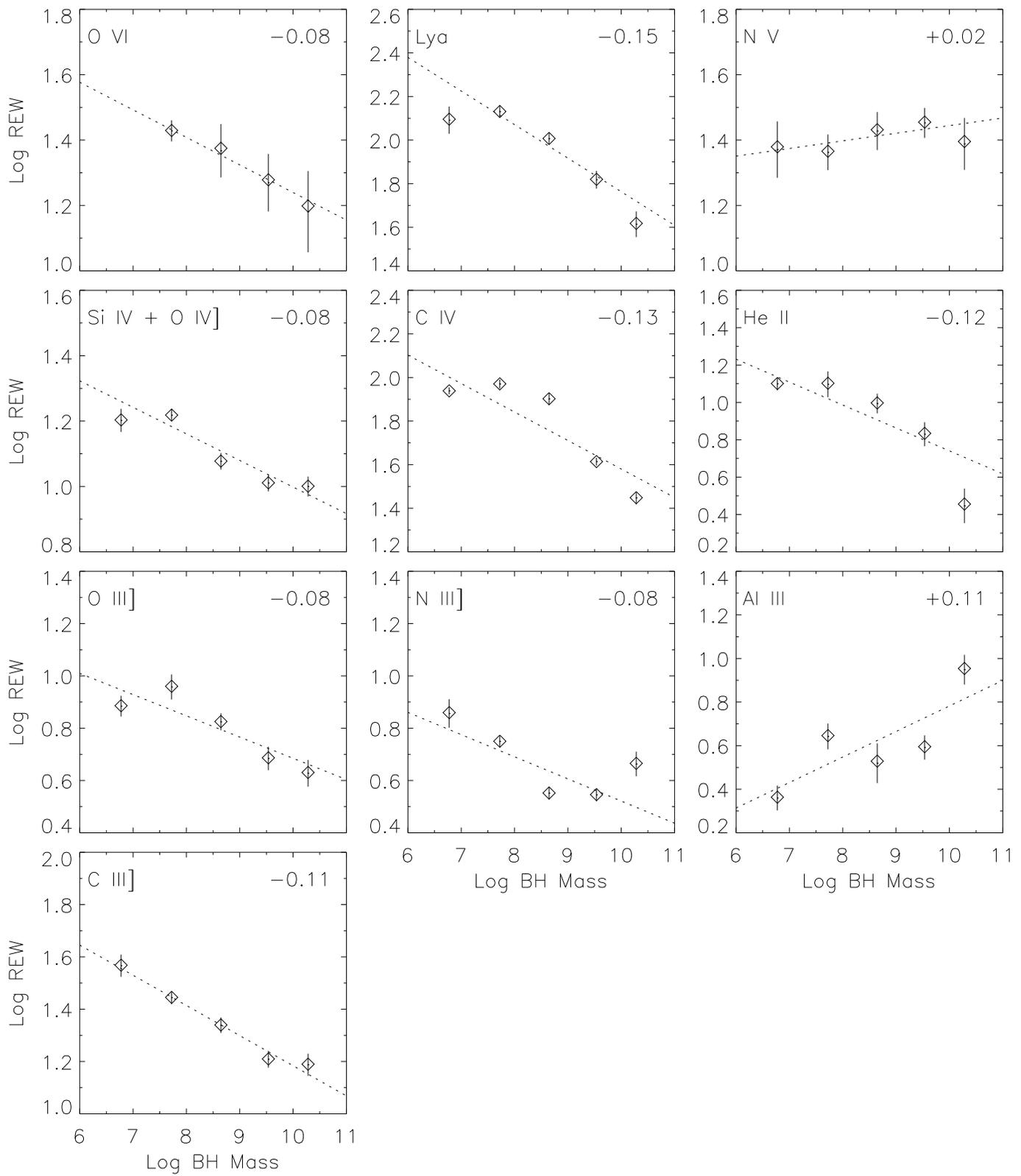}
}
\vskip 14pt
\vskip 14pt
\caption{
Emission line equivalent widths as a function of SMBH mass.  We calculated
linear fits to the EWs for the entire SMBH mass range using a chi-square
minimization routine (dotted lines).  The slopes of these fits are given in
the upper right hand corner of each plot.}
}
\end{figure}

\begin{figure}
\vbox{
\centerline{
\psfig{figure=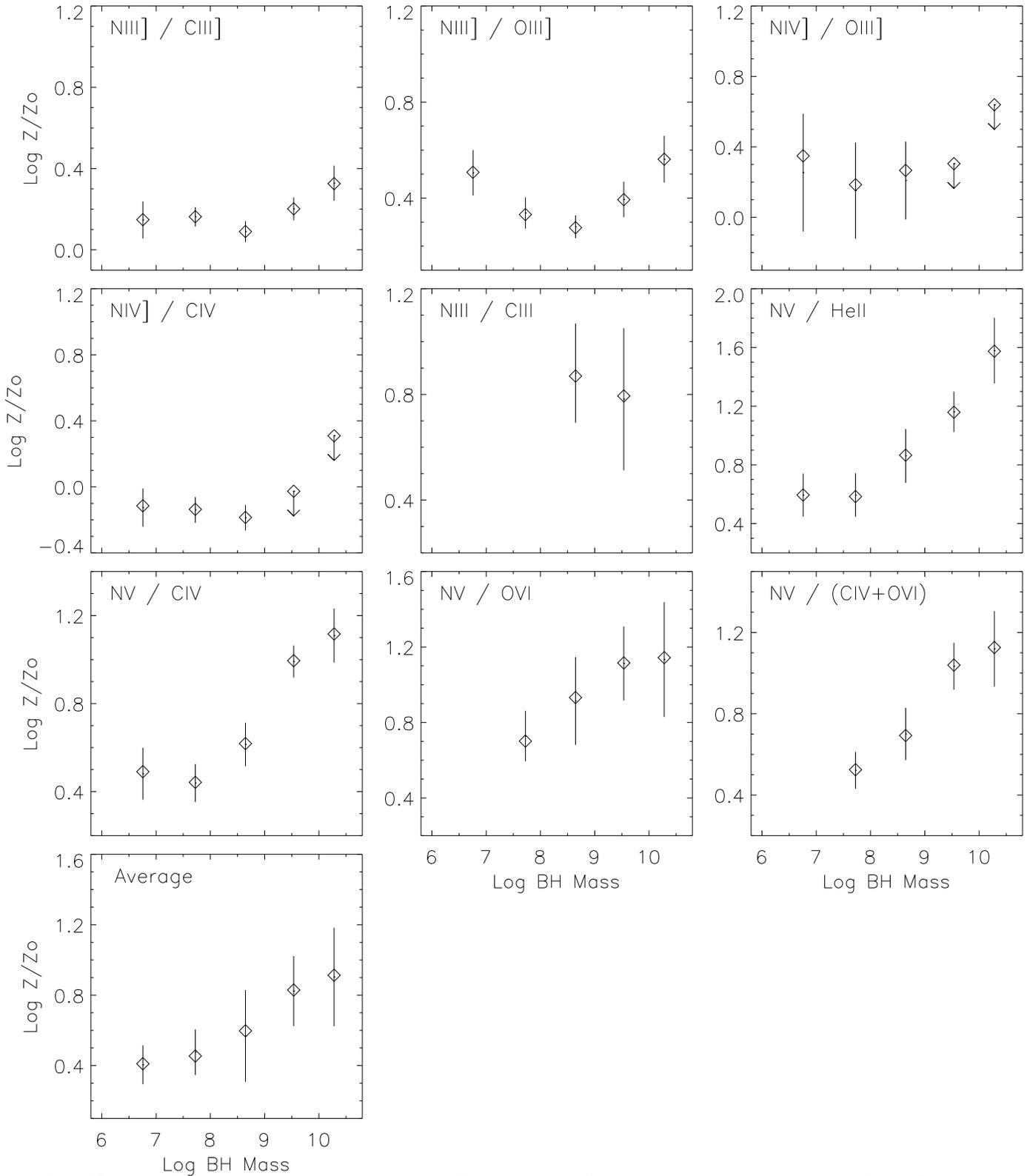}
}
\vskip 14pt
\caption
{
Metallicities derived from comparisons of different line ratio diagnostics
to theoretical results from Figure 5 in Hamann et al. (2002)
are shown as a function of SMBH mass.  The average includes
\ion{N}{3}]/\ion{C}{3}], \ion{N}{3}]/\ion{O}{3}], \ion{N}{5}/\ion{C}{4},
and \ion{N}{5}/\ion{O}{6}.} 
}
\end{figure}

\begin{figure}
\vbox{
\centerline{
\psfig{figure=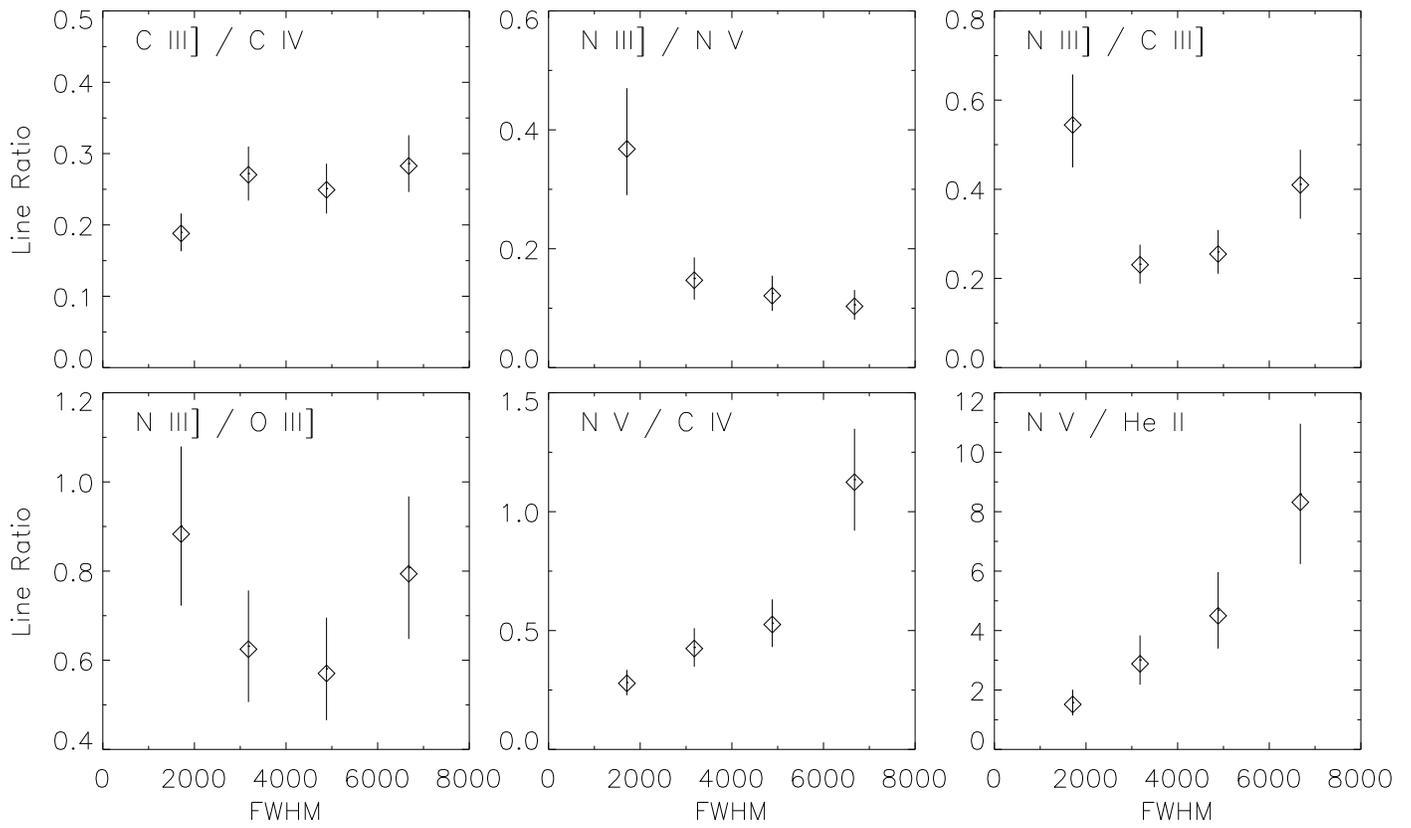}
}
\vskip 14pt
\caption
{Emission line ratios are shown as a function of the FWHM of \ion{C}{4}.
}
}
\end{figure}

\begin{deluxetable}{cccccccc}
\tabletypesize{\small}
\tablecaption{Composite Parameters}
%\tablewidth{535.22502pt}
\tablewidth{485.22502pt}
\tablehead{
\colhead{Log $M_{\rm SMBH}$} &
\colhead{$\alpha$} &
\colhead{\# Objects} &
\colhead{FWHM(\ion{C}{4})} &
\colhead{\# Objects} &
\colhead{FWHM($H \beta$)} &
\colhead{Log $\lambda L_{\lambda}(1450 {\rm \AA})$} &
\colhead{Log $M_{\rm mean}$} \\
\colhead{[\Msun]} &
\colhead{} &
\colhead{at \ion{C}{4}} &
\colhead{[km s$^{-1}$]} &
\colhead{at $H \beta$} &
\colhead{[km s$^{-1}$]} &
\colhead{[ergs s$^{-1}$]} &
\colhead{[\Msun]}
}
\tablecolumns{8}
\startdata
$6-7$ & -0.98 & 7 & 2100 & 4 & 1400 & 44.0 & 6.78\\
\noalign{\vskip 4pt}
$7-8$ & -0.90 & 61 & 3300 & 31 & 2250 & 45.0 & 7.27\\
\noalign{\vskip 4pt}
$8-9$ & -0.71 & 198 & 3900 & 35 & 3000 & 46.2 & 8.65\\
\noalign{\vskip 4pt}
$9-10$ & -0.59 & 261 & 4900 & 6 & 3650 & 46.9 & 9.54\\
\noalign{\vskip 4pt}
$>10$ & -0.57 & 34 & 6500 & 1 & 4500 & 47.5 & 10.28\\
\enddata
\end{deluxetable}

\begin{deluxetable}{lccccccccccccccc}
\rotate
\tabletypesize{\scriptsize}
%\tabletypesize{\tiny}
\tablecaption{Emission Line Data}
\tablewidth{685.22502pt}
%\tablewidth{645.22502pt}
\tablehead{
\colhead{} &
\multicolumn{3}{c}{-------- $10^{6} \Msun$ --------} &
\multicolumn{3}{c}{-------- $10^{7} \Msun$ --------} &
\multicolumn{3}{c}{-------- $10^{8} \Msun$ --------} &
\multicolumn{3}{c}{-------- $10^{9} \Msun$ --------} &
\multicolumn{3}{c}{-------- $10^{10} \Msun$ --------} \\ 
\colhead{Line} &
\colhead{Flux/$\Lya$} &
\colhead{REW${^a}$} &
\colhead{FWHM$^{b}$} &
\colhead{Flux/$\Lya$} &
\colhead{REW${^a}$} &
\colhead{FWHM$^{b}$} &
\colhead{Flux/$\Lya$} &
\colhead{REW${^a}$} &
\colhead{FWHM$^{b}$} &
\colhead{Flux/$\Lya$} &
\colhead{REW${^a}$} &
\colhead{FWHM$^{b}$} &
\colhead{Flux/$\Lya$} &
\colhead{REW${^a}$} &
\colhead{FWHM$^{b}$}
}
\tablecolumns{16}
\startdata
\ion{C}{3} $\lambda 977$ & -- & -- & -- & -- & -- & --
& 0.023 & 2 & 3830 & 0.027 & 2 & 3170 & -- & -- & --\\
\ion{N}{3} $\lambda 991$ & -- & -- & -- & -- & -- & --
& 0.029 & 3 & 3830 & 0.030 & 2 & 3170 & 0.057 & 2 & 3350\\
\ion{O}{6} $\lambda 1035$ & -- & -- & -- & 0.207 & 27 & 5800
& 0.242 & 24 & 4900 & 0.299 & 19 & 5700 & 0.396 & 16 & 8600\\
$\Lya$ $\lambda 1216$ & 1.000 & 125 & 1500 & 1.000 & 135 & 2500
& 1.000 & 102 & 3000 & 1.000 & 66 & 3800 & 1.000 & 42 & 6700\\
\ion{N}{5} $\lambda 1240$ & 0.191 & 24 & 2100 & 0.171 & 23 & 3100
& 0.264 & 27 & 3800 & 0.430 & 29 & 4700 & 0.597 & 25 & 6500\\
\ion{Si}{2} $\lambda 1263$ & 0.010 & 1 & 3070 & 0.012 & 2 & 3860
& 0.022 & 2 & 5010 & 0.026 & 2 & 4450 & 0.047 & 2 & 5340\\
\ion{O}{1} $\lambda 1303$ & 0.027 & 4 & 2000 & 0.027 & 4 & 4110
& 0.044 & 5 & 4650 & 0.052 & 4 & 4600 & 0.087 & 4 & 6320\\
\ion{C}{2} $\lambda 1335$ & 0.029 & 4 & 4140 & 0.015 & 2 & 3170
& 0.019 & 2 & 5380 & 0.027 & 2 & 4290 & 0.038 & 2 & 4360\\
\ion{Si}{4} + \ion{O}{4}] & 0.113 & 16 & 4240 & 0.108 & 17 & 5340
& 0.102 & 12 & 5890 & 0.133 & 11 & 6340 & 0.205 & 10 & 7250\\
\ion{N}{4}] $\lambda 1486$ & 0.019 & 3 & 2100 & 0.015 & 2 & 3100
& 0.014 & 2 & 3800 & 0.016$^{u}$ & 1$^{u}$ & 4700
& 0.033$^{u}$ & 2$^{u}$ & 6500\\
%\ion{Si}{2} $\lambda 1531$ & 0.004 & 1 & 3070 & 0.007 & 1 & 3640
%& 0.008 & 1 & 5010 & 0.007 & 1 & 4450 & -- & -- & --\\
\ion{C}{4}  $\lambda 1549$ & 0.556 & 87 & 2100 & 0.547 & 94 & 3100
& 0.597 & 80 & 3800 & 0.462 & 41 & 4700 & 0.500 & 28 & 6500\\
\ion{He}{2} $\lambda 1640$ & 0.076 & 13 & 2300 & 0.070 & 13 & 4200
& 0.069 & 10 & 4300 & 0.071 & 7 & 6500 & 0.047 & 3 & 6800\\
\ion{O}{3}] $\lambda 1665$ & 0.046 & 8 & 2000 & 0.049 & 9 & 3200
& 0.046 & 7 & 3800 & 0.049 & 5 & 5000 & 0.068 & 4 & 6300\\
\ion{N}{3}] $\lambda 1751$ & 0.041 & 7 & 2100 & 0.029 & 6 & 3100
& 0.023 & 4 & 3800 & 0.033 & 4 & 4700 & 0.069 & 5 & 6500\\
\ion{Al}{3} $\lambda 1859$ & 0.012 & 2 & 2050 & 0.021 & 4 & 3460
& 0.020 & 3 & 3630 & 0.034 & 4 & 4430 & 0.122 & 9 & 6690\\
\ion{Si}{3}] $\lambda 1892$ & 0.023 & 4 & 2050 & 0.007 & 1 & 3460
& 0.007 & 1 & 3630 & 0.008 & 1 & 4430 & 0.010 & 1 & 6690\\
\ion{C}{3}] $\lambda 1909$ & 0.192 & 37 & 2300 & 0.130 & 28 & 3800
& 0.125 & 22 & 4200 & 0.135 & 16 & 6800 & 0.203 & 16 & 6700\\
\enddata
\tablenotetext{a}{In units of \AA}
\tablenotetext{b}{In units of km~s$^{-1}$}
\tablenotetext{u}{Upper limit for \ion{N}{4}}
\end{deluxetable}

\end{document}